\documentclass{article}
\usepackage[english]{babel} 
\usepackage{mathdots}
\usepackage{amsmath}
\usepackage{amssymb}
\usepackage{graphicx}
\usepackage{epstopdf}
\usepackage{inputenc}
\usepackage{geometry}
\usepackage{authblk}
\usepackage{hyperref}
\providecommand{\keywords}[1]{\textbf{\textit{Key Words and Phrases:}} #1}

\begin{document}
	
	\title{%
		\textbf{Exact solutions for a two-dimensional thermoelectric MHD flow with steady-state heat transfer on a vertical plate with two instantaneous infinite hot suction lines}  \\
		\large \it{This work is dedicated to the memory of our professor ``Ibrahim H. El-Sirafy'' }}
	
	\author[1,3]{\textbf{Emad Awad}\thanks{emadawad78@alexu.edu.eg}}
	\author[2]{\textbf{Sayed M. Abo-Dahab}\thanks{sdahb@yahoo.com}}
	\author[1]{\textbf{Mohamed A. Abdou }\thanks{abdella\_777@yahoo.com}}
	\affil[1]{Department of Mathematics, Faculty of Education, Alexandria University, Souter St. El-Shatby, Alexandria P.O. box 21526, Egypt.}
	\affil[2]{Department of Mathematics, Faculty of Science, South Valley University, Qena 83523, Egypt}
	\affil[3]{Author to whom any correspondence should be addressed}
	\renewcommand\Authands{ and }

	\maketitle

\begin{abstract}
	 In the present work, the mathematical description of a two-dimensional unsteady magneto-hydrodynamics slow flow with thermoelectric properties (TEMHD) on an infinite vertical partially hot porous plate is presented. The Laplace-Fourier transform technique is employed to simplify the field equations. For the steady-state heat transfer assumption, exact expressions for the temperature and stream function are analytically obtained for TEMHD flow on a wall with two infinite instantaneous hot suction lines. The results obtained are displayed graphically.  
\end{abstract}
\noindent 


\tableofcontents

\section{ Introduction}

Liquid metals have shown their efficiency in cooling the fusion machines when used as the inner surface material instead of solid materials (such as tungsten and graphite) which are conventionally used as plasma facing components (PFCs) \cite{r1}. It is believed that the thermoelectric currents arisen in a liquid metal in contact with a solid boundary of markedly distinct thermopower and interfacial temperature distribution are capable to dissipate the heat by transferring thermal energy from the hot spot to the cold region. The basics of TEMHD flow were presented by Shercliff \cite{r2}, who also followed his work with a triple of investigations including TEMHD flow in closed containers \cite{r3}, in contact with a wall \cite{r4}, and the pipe end problem \cite{r5}. Recent experiments, Solid/Liquid Lithium Divertor Experiment (SLIDE), have been conducted at University of Illinois at Urbana-Champaign \cite{r6,r7} have reported the direct observation of TEMHD driven flow, and developed the Lithium/Metal Infused Trenches (LiMITs) concept. 

Because of the complexity of modeling such a type of problems, the numerical techniques, such as time series approximations using FORTRAN for the one-dimensional simulation \cite{r8, r9}, the finite element method \cite{r10}, or the finite volume method \cite{r11}, are indispensable, to name a few. However, the analytical expressions provided in \cite{r2,r3,r4,r5} still show a full resolution even though the time variable is absent.  Since additional dimensions in the space will further complicate the mathematics, finding analytical solutions requires certain restrictions on the fluid motion.

Prandtl number ($p_r$) is an intrinsic property of the fluid motion expressing the ratio of momentum diffusivity (kinematic viscosity) to the thermal diffusivity. The active materials in Shercliff studies, liquid metals, often have Prandtl numbers ranging from $0.001$ to $0.01$, (see, p. 1005 in Ref. \cite{r12}). On the other hand, the numerical work due to Thual  \cite{r13} elucidates that the properties of low-$p_r$ liquids are quite close to those with zero-$p_r$ in the no-slip boundary condition, see also \cite{r14}. In the present study, we shall obtain explicit, calculable integral expressions for the solution of two-dimensional (2D), unsteady, slow TEMHD flow with Prandtl numbers close to zero such that the heat transfer can be considered in the steady state. We shall disregard the turbulent motion term and the natural convection process in the zero-$p_r$ ($p_r=0$).

The MHD flow past a vertical plate was the subject of various recent studies. For example, Veera Krishna with coworkers \cite{r15,r16,r17} have studied the effects of thermodiffusion, chemical reaction, Soret, Joule and Hall effects on the MHD flow past a porous vertical plate, see also \cite{r18}. Dadheech \textit{et al}.  \cite{r19} introduced a detailed analysis for MHD Williamson fluid slip-flow over a vertical plate with Cattaneo-Christov heat flux. In the recent case study \cite{r20}, the authors have studied the chemical reaction effect on the MHD micropolar fluid flow past over a vertical Riga plate. Another set of studies reports the possibilities of obtaining exact formulas for the solution of slow flow problems, (see El-Sirafy and coworkers \cite{r21,r22,r23,r24,r25}). 

In this article, we consider the 2D slow TEMHD flow of Liquid-Metal contiguous with a vertical infinite partially porous plate with a hot portion in the presence of a vertical uniform magnetic field. The induced electric current will be eliminated between the Fourier and Ohm laws, and the induced electric field will be neglected compared with the fluid motion so that the Maxwell's equations and the electrical boundary conditions will disappear. The Green's functions for the temperature and the stream function are obtained firstly, then, the exact solutions for 2D unsteady TEMHD slow flow on a wall with two infinite instantaneous hot suction lines are provided. The paper is organized as follows: the problem is formulated, and the field equations are simplified using the Laplace-Fourier transform in section \ref{sec2}. Exact solutions for 2D TEHMD, in the case of steady state heat flow, are derived in section  \ref{sec3}. In section  \ref{sec4}, graphical representations and discussions about the derived expressions for temperature, stream function and velocity are represented. Finally, we draw our conclusions in section  \ref{sec5}.

\section{ Model}
\label{sec2}

\subsection{ Field equations}

The theory of thermoelectric MHD was firstly presented by Shercliff \cite{r2} wherein the classical Ohm and Fourier laws are replaced by the generalized constitutive laws, see also \cite{r8, r26, r27}:

\begin{subequations}
	\begin{equation}\label{1a}
		 J_i={\sigma }_0\left[E_i+{\left({\mathbf{v}}\times {\mathbf{B}}\right)}_i-k_0T_{,i}\right],
	\end{equation}
    \begin{equation}\label{1b}
	Q_i=-\kappa T_{,i}+{\pi }_0J_i,
    \end{equation}
\end{subequations}
where $J_i$, $E_i$ and $B_i$ are respectively the ith component of the induced electric current\textbf{ }${\mathbf{J}}$, induced electric density ${\mathbf{E}}$ and magnetic flux density ${\mathbf{B}}$, $\mathbf{v}=\left\langle u,v,\omega \right\rangle $ is the velocity vector with ith component $v_i$, $T$ is the absolute temperature and $Q_i$ is the component of the heat flux vector ${\mathbf{Q}}$ in $x_i$ direction. Moreover, ${\sigma }_0$ stands for the electric conductivity, $k_0$ is the Seebeck coefficient (thermoelectric power), ${\pi }_0$ is the Peltier coefficient, $\kappa $ is the thermal conductivity. From now on we will consider constant material parameters, namely, ${\sigma }_0$, $k_0$, $\kappa $ and ${\pi }_0$ are constants. The energy balance equation for incompressible flow is given as \cite{r28}:

\begin{equation}\label{2}
\rho c_p\frac{DT}{Dt}=-Q_{i,i}+\mathrm{\Phi },
\end{equation}
where $c_p$ is the specific heat capacity at constant pressure, $D/Dt$ is defined through

\begin{equation}\label{3}
\frac{D}{Dt}=\frac{\partial }{\partial t}+\left({\mathbf{v}}\cdot \mathbf{\nabla}\right),
\end{equation}
$\mathbf{\nabla}$ is the Laplace operator, and $\mathrm{\Phi }$ is known as the internal heat due to viscous stresses and given in terms of the cartesian coordinates as 

\begin{equation}\label{4}
 \mathrm{\Phi }\mathrm{=}\mu \left[2{\left(\frac{\partial u}{\partial x}\right)}^2+2{\left(\frac{\partial v}{\partial y}\right)}^2+2{\left(\frac{\partial \omega }{\partial z}\right)}^2+{\left(\frac{\partial \omega }{\partial y}+\frac{\partial v}{\partial z}\right)}^2+{\left(\frac{\partial u}{\partial z}+\frac{\partial \omega }{\partial x}\right)}^2+{\left(\frac{\partial u}{\partial y}+\frac{\partial v}{\partial x}\right)}^2\right],
\end{equation}
where $x,y$ and $z$ are the Cartesian coordinates, and $\mu $ is the dynamic viscosity. In Eq. \eqref{2}, $\rho $ is the fluid density satisfying the continuity condition $\partial \rho /\partial t+\mathbf{\nabla}\cdot \left(\rho {\mathbf{v}}\right)=0$, which reduces to the form

\begin{equation}\label{5}
\mathbf{\nabla}\cdot {\mathbf{v}}=0,
\end{equation}
in the case of constant density. Lastly, the equation of motion for MHD flow is given by

\begin{equation}\label{6}
\rho \frac{Dv_i}{Dt}={\left({\mathbf{J}}\times {\mathbf{B}}\right)}_i+{\sigma }_{ij,j},
\end{equation}
where ${\sigma }_{ij}$ is the stress tensor determined through

\begin{equation}\label{7}
 {\sigma }_{ij}=-P{\delta }_{ij}+\mu \left(v_{i,j}+v_{j,i}\right),
\end{equation}
where $P$ is the pressure and ${\delta }_{ij}$ is the Kronecker delta. The performance of thermoelectric Liquid depends on the characteristic material parameter, $Z$, known as the figure of merit and well-known through the relation

\begin{equation}\label{8}
Z\left(T\right)=\frac{{\sigma }_0k^2_0}{\kappa }=\frac{{\sigma }_0{\pi }_0k_0}{\kappa T}.
\end{equation}

In \eqref{8}, the relation between Seebeck and Peltier coefficients is utilized, i.e., ${\pi }_0=k_0T$. Generally speaking, the figure of merit is a temperature-dependent property, see e.g., in \cite{r27}, however, in the presence of constant material parameters, all the material parameters are evaluated at the room temperature $T_{\infty }$, then we can define the figure of merit in the following form:

\begin{equation}\label{9}
ZT_{\infty }=Z\times T_{\infty }=\frac{{\sigma }_0{\pi }_0k_0}{\kappa }.
\end{equation}

The induced electromagnetic fields satisfy the Maxwell equations \cite{r26, r27} without induced electric charge:

\begin{equation}\label{10}
\mathbf{\nabla}\times {\mathbf{H}}={\mathbf{J}}+{\varepsilon }_0\frac{\partial {\mathbf{E}}}{\partial t},\ \ \ \mathbf{\nabla}\times {\mathbf{E}}=-{\mu }_0\frac{\partial {\mathbf{H}}}{\partial t},\ \ \ \mathbf{\nabla}\cdot{\mathbf{H}}=0,\ \ \ \mathbf{\nabla}\cdot{\mathbf{E}}=0,
\end{equation}
where ${\varepsilon }_0$ is the vacuum permittivity and ${\mu }_0$ is the magnetic permeability. In the case of slow motion, we replace the operator \eqref{3} by the direct temporal differential operator:

\begin{equation}\label{11}
\frac{D}{Dt}=\frac{\partial }{\partial t},
\end{equation}
 and the viscous term \eqref{4} is neglected.  

\subsection{ Slow TEMHD viscous flow on a wall}

\begin{figure}
	\centering
	\includegraphics*[scale=0.4,angle=0]{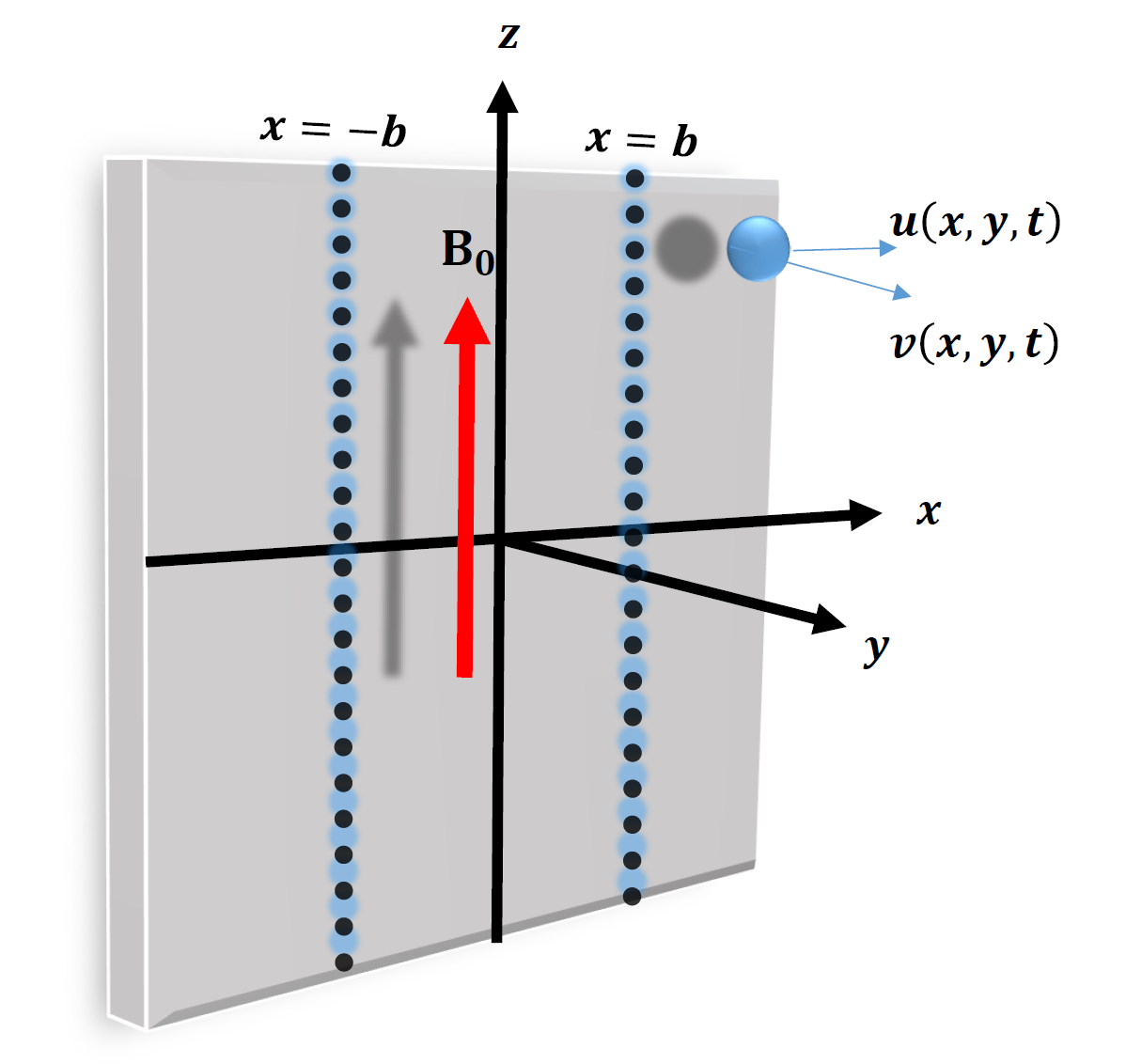}
	\caption{Physical description of the problem.}
	\label{fig1} 
\end{figure}

We consider the laminar two-dimensional slow flow of a viscous incompressible fluid occupying the region $\mathrm{\Omega }=\left\{y>0,\ \left.-\infty <x,z<\infty |t>0\right\}\right.$, in contact with an infinite vertical plate occupying the $xz$-plane where $z$-axis is taken vertically upwards, and $y$-axis is in the normal direction. A uniform magnetic field is applied in the vertical direction ${{\mathbf{H}}}_0=\left\langle 0,0,H_0\right\rangle $. In order to make the problem more tractable, we assume that ${{\mathbf{H}}}_0$ will not be distorted, namely, ${\mathbf{H}}={{\mathbf{H}}}_0$ and ${\mathbf{E}}=0$, see \cite{r29}. A hot porous portion of the plate with an infinite length ($-\infty <z<\infty $), is distinguished from the rest of this plate, and is the cause for the two-dimensional slow flow (in our case, two vertical suction hot lines). Let $T_w$ be the temperature of the hot porous portion, and $T_{\infty }$ be the temperature of the other part of the vertical plate and the temperature at infinity. Therefore, the TEMHD flow will be induced by the contact with a different thermoelectric material, the interfacial temperature distribution, and the applied undistorted uniform magnetic field ${{\mathbf{H}}}_0$. As described in Fig. \ref{fig1}, the induced electric current is perpendicular to the magnetic induction field, i.e.,  $J=\left\langle J_x\left(x,y,t\right),{\ J}_y\left(x,y,t\right),0\right\rangle $ with Lorentz force ${\mathbf{J}}\times {{\mathbf{B}}}_{0}$ in $xy$-plane, where ${{\mathbf{B}}}_{0}={\mu }_0\mathrm{\ }{{\mathbf{H}}}_0=\left\langle 0,0,B_0\right\rangle $ is the applied magnetic induction vector (or flux density). The fluid motion is described as ${\mathbf{v}}=\left\langle u\left(x,y,t\right),v\left(x,y,t\right),0\right\rangle $. Therefore, the governing equations of the two-dimensional slow MHD viscous flow with thermoelectric properties are given by

\begin{subequations}\label{12}
	\begin{equation}\label{12a}
		{\partial }_xu+{\partial }_yv=0,
	\end{equation}
    \begin{equation}\label{12b}
    	\left({\mathbf{\nabla}}^{\mathrm{2}}_p-{\partial }_t-M\right)u={\partial }_xP+k_1{\partial }_y\theta ,
    \end{equation}
    \begin{equation}\label{12c}
    	\left({\mathbf{\nabla}}^{\mathrm{2}}_p-{\partial }_t-M\right)v={\partial }_yP-k_1{\partial }_x\theta ,
    \end{equation}
    \begin{equation}\label{12d}
    	\left(a{\mathbf{\nabla}}^{\mathrm{2}}_p-{p_r\partial }_t\right)\theta =k_1k_2\left({\partial }_xv-{\partial }_yu\right),
    	\end{equation}
\end{subequations}
where ${\partial }_{\varsigma }$ denotes a differentiation with respect to the variable $\varsigma $, and ${\mathbf{\nabla}}^{\mathrm{2}}_p\mathrm{=}{\mathrm{\partial }}^{\mathrm{2}}_x\mathrm{+}{\mathrm{\partial }}^{\mathrm{2}}_y$ is the Laplace operator defined on $xy$-plane,

\begin{align}\label{13}
M=\frac{R_H}{P_m},\quad P_m=&\frac{1}{{\sigma }_0{\mu }_0\nu },\ \ \ R_H=\frac{{\mu }_0H^2_0}{\rho V^2_0},\ \ \ k_1=\frac{{\sigma }_0B_0k_0T_d}{\rho V^2_0},\ \ \ k_2=\frac{\mu V^2_0T_{\infty }}{\kappa T^2_d}, \nonumber \\
p_r=&\frac{\mu c_p}{\kappa },\ \ \ a=1+ZT_{\infty },\ \ \ {\pi }_0=k_0T_{\infty },
\end{align}
$ZT_{\infty }$ is given by \eqref{9}, $P_m$ is the magnetic Prandtl number \cite{r30}, and $R_H$ is the magnetic pressure number \cite{r31}. The dimensionless stresses are given by

\begin{equation}\label{14}
{\sigma }_{xx}=-P+2{\partial }_xu,\ \ \ {\sigma }_{yy}=-P+2{\partial }_yv,\ \ \ {\sigma }_{xy}={\partial }_xv+{\partial }_yu.
\end{equation}

It is noteworthy to mention that in adapting equations \eqref{12} and \eqref{14}, we have employed the dimensionless transformations:

\begin{align}\label{15}
x\to \frac{\nu }{V_0}x,\ \ \ y\to \frac{\nu }{V_0}y,\ \ \ &t\to \frac{\nu }{V^2_0}t,\ \ \ u\to V_0u,\ \ \ v\to V_0v,\ \ \ T\to T_{\infty }+T_d\theta ,\nonumber \\
&P\to \rho V^2_0P,\ \ \ {\sigma }_{ij}\to \rho V^2_0{\sigma }_{ij},
\end{align}
where $V_0$ is a characteristic velocity constant, $\nu =\mu /\rho $ is the kinematic viscosity, and $T_d=T_w-T_{\infty }$. Finally, we assume that the vertical plate (wall) is partially porous with suction velocity, and the surface temperature is prescribed on the wall, $y=0$. As such, the following set of initial and boundary conditions will be appended to the above governing equations:

\begin{subequations}\label{16}
	\begin{equation}\label{16a}
		u\left(x,y,0\right)=v\left(x,y,0\right)=\theta \left(x,y,0\right)=0,\ \ \ y>0,
	\end{equation}
    \begin{equation}\label{16b}
    	\frac{\partial u\left(x,0,t\right)}{\partial y}=0,\ \ \ v\left(x,0,t\right)=U_0\left(x,t\right),\ \ \ \theta \left(x,0,t\right)=T_0\left(x,t\right),\ \ \ t\ge 0,
    \end{equation}
      \begin{equation}\label{16c}
     {\left.u\left(x,y,t\right)\right|}_{y\to +\infty }\to 0,\ \ \ {\left.v\left(x,y,t\right)\right|}_{y\to +\infty }\to 0,\ \ \ {\left.\theta \left(x,y,t\right)\right|}_{y\to +\infty }\to 0,
     \end{equation}
\end{subequations}
where the first condition of \eqref{16b}, representing the mathematical form of no-slip boundary condition, i.e., $u_{liquid}-u_{wall}=\beta {\left.\partial u/\partial y\right|}_{wall}\ $, $\beta $ is the slip length, was considered previously in viscous flow \cite{r32}, the conditions in Eq. \eqref{16c} lead to bounded solutions at infinity, and $U_0\left(x,t\right)$ and $T_0\left(x,t\right)$ are prescribed functions.

In accordance with the conservation of mass Eq. \eqref{12a}, we can assume the potential representation ${\mathbf{v}}=\mathbf{\nabla}\times {\mathbf{\Psi }}$, where ${\mathbf{\Psi }}$ is a vector potential with the form ${\mathbf{\Upsilon }}=\psi {{\mathbf{e}}}_{\mathrm{3}}$ and ${{\mathbf{e}}}_{\mathrm{3}}$ is the unit along $z$-axis, thereby we introduce the following velocity representation \cite{r28, r32, r33}

\begin{equation}\label{17}
u\left(x,y,t\right)={\partial }_y\psi \left(x,y,t\right),\ \ \ v\left(x,y,t\right)=-{\partial }_x\psi \left(x,y,t\right). 
\end{equation}

The assumption \eqref{17} transforms the field equations \eqref{12} subject to \eqref{16}, to the forms

\begin{equation}\label{18}
 \left({\mathbf{\nabla}}^{2}_p-{\partial }_t-M\right){\mathbf{\nabla}}^{2}_p\psi =k_1{\mathbf{\nabla}}^{\mathrm{2}}_p\theta ,
\end{equation}
\begin{equation}\label{19}
\left(a{\mathbf{\nabla}}^{2}_p-{p_r\partial }_t\right)\theta =-k_1k_2{\mathbf{\nabla}}^{2}_p\psi ,
\end{equation}
subject to the initial and boundary conditions

\begin{equation}\label{20}
\psi \left(x,y,0\right)=\theta \left(x,y,0\right)=0,\ \ \ y>0,
\end{equation}
\begin{subequations}\label{21}
	\begin{equation}\label{21a}
		{\partial }^2_y\psi \left(x,0,t\right)=0,\ \ \ {\partial }_x\psi \left(x,0,t\right)=-U_0\left(x,t\right),\ \ \ \theta \left(x,0,t\right)=T_0\left(x,t\right).
	\end{equation}
\begin{equation}\label{21b}
	{\left.\psi \left(x,y,t\right)\right|}_{y\to +\infty }\to 0,\ \ \ {\left.\theta \left(x,y,t\right)\right|}_{y\to +\infty }\to 0,
\end{equation}
\end{subequations}

\subsection{ Steady-state heat flow }

In the liquid metal motion, the kinematic viscosity is much smaller than the thermal diffusivity \cite{r12}, i.e., $\nu \ll {\alpha }_T$, and ${\alpha }_T=\kappa /\left(\rho c_p\right)$. In this section, we provide a trial solution for the case of steady-state heat transfer, or alternatively, the limit of low Prandtl number when compared to time, in which the energy balance equation \eqref{19} reduces to, \cite{r13}:

\begin{equation}\label{22}
a{\mathbf{\nabla}}^{\mathrm{2}}_p\theta =-k_1k_2{\mathbf{\nabla}}^{\mathrm{2}}_p\psi .
\end{equation}

In \eqref{22}, we neglect the free convection term resulting from the vertical velocity component. It is noted that the presence of thermoelectric term in the right side of \eqref{22} was not considered by Thual, \textit{cf}. relation (2.3) in [13]. 

Utilizing the Laplace-Fourier transform technique \footnote{The tildes refer to the Laplace transform defined for any generic function as $\tilde{f}\left(x,y,s\right)=\mathcal{L}\left\{f\left(x,y,t\right);t\right\}(x,y,s)=\int^{\infty }_0{f\left(x,y,t\right)\exp(-st)}dt$ and the hats refer to the Fourier transform defined as $\hat{f}\left(q,y,t\right)=\mathcal{F}\left\{f\left(x,y,t;x\right)\right\}(q,y,t)=\int^{\infty }_{-\infty }{f\left(x,y,t\right){\exp\left(-iqx\right)}dx}$,where $s\in \mathbb{C}$ is the Laplace parameter and $q\in \mathbb{R}$ is the Fourier parameter.}, we find the following trial solution for zero-Prandtl number MHD flow governed by Eqs \eqref{18}-\eqref{19} and \eqref{22} subject to the initial and boundary conditions \eqref{20}-\eqref{21}:

\begin{align}\label{23}
\widehat{\widetilde{\psi }}\left(q,y,s\right)=A_1\left(q,s\right)e^{-r_1y}+A_2\left(q,s\right)e^{-r_2y}, \nonumber \\
\widehat{\widetilde{\theta }}\left(q,y,s\right)=B_1\left(q,s\right)e^{-r_1y}+B_2\left(q,s\right)e^{-r_2y},
\end{align}
where $A_1$, $A_2$, $B_1$ and $B_2$, are unknown functions of $q$ and $s$ to be determined, the boundary conditions \eqref{21b} are considered, and $r_1$ and $r_2$ are the roots of the characteristic equation:
\[\left(r^2-q^2\right)\left[r^2-\left(q^2+s+M-\frac{k^2_1k_2}{a}\right)\right]=0.\] 
Namely, 

\begin{equation}\label{24}
r_1=\left|q\right|,\ \ r_2=\sqrt{q^2+s+\frac{M}{a}},
\end{equation}
where we have utilized the relations $k^2_1k_2=MZT_{\infty }$ and $a=1+ZT_{\infty }$. Moreover, use of the trial solution \eqref{23} and the energy balance equation \eqref{22} reduces the number of parameters, and then the trial solutions read

\begin{align}\label{25}
\widehat{\widetilde{\psi }}\left(q,y,s\right)=A_1\left(q,s\right)e^{-r_1y}+A_2\left(q,s\right)e^{-r_2y}, \nonumber \\
\widehat{\widetilde{\theta }}\left(q,y,s\right)=B_1\left(q,s\right)e^{-r_1y}-\frac{k_1k_2}{a}A_2\left(q,s\right)e^{-r_2y},
\end{align}
where $r_1$ and $r_2$ are given by \eqref{24}.

\section{ Exact Results}
\label{sec3}

In this section, we derive the Green's functions for  the temperature and stream function for TEMHD with steady-state heat flow, given in the Laplace-Fourier domain through equation \eqref{25}, see similar illustrative examples in the mass diffusion on the infinite domain \cite{r34,r35,r36,r37} and in heat transfer problems \cite{r38}. With the help of boundary conditions on the wall, Eq. \eqref{21a}, the trial solution \eqref{25} can be written as

\begin{align}\label{26}
\widehat{\widetilde{\psi}}\left(q,y,s\right)=iq\frac{{\widehat{\tilde{U}}}_0\left(q,s\right)}{s+\frac{M}{a}}\left(e^{-r_1y}-e^{-r_2y}\right)+\frac{i}{q}{\widehat{\tilde{U}}}_0\left(q,s\right)e^{-r_1y},\nonumber \\ 
\widehat{\widetilde{\theta}}\left(q,y,s\right)=iq\frac{k_1k_2{\widehat{\tilde{U}}}_0\left(q,s\right)}{a\left(s+\frac{M}{a}\right)}\left(e^{-r_2y}-e^{-r_1y}\right)+{\widehat{\tilde{T}}}_0\left(q,s\right)e^{-r_1y}.
\end{align}
 Inverting the Fourier and Laplace transforms, we get the integral representations:

\begin{align}\label{27}
	\psi \left(x,y,t\right)=\frac{2y}{\pi }{\exp \left(-\frac{M}{a}t\right)}&\int^t_0{\int^{\infty }_{-\infty }{\frac{\left(x-\xi \right)U_0\left(\xi ,\tau \right){\exp \left(\frac{M}{a}\tau \right)}}{{\left[{\left(x-\xi \right)}^2+y^2\right]}^2}}}\left[\mathrm{\Gamma }\left(2,\frac{{\left(x-\xi \right)}^2+y^2}{4\left(t-\tau \right)}\right)-1\right]d\xi d\tau \nonumber \\
&-\frac{1}{\pi }\int^{\infty }_{-\infty }{U_0\left(\xi ,\tau \right){{\mathrm{tan}}^{-1} \left(\frac{x-\xi }{y}\right)}d\xi },
\end{align}

\begin{align}\label{28}
	\theta \left(x,y,t\right)=\frac{2k_1k_2y}{\pi a}{\exp \left(-\frac{M}{a}t\right)}&\int^t_0{\int^{\infty }_{-\infty }{\frac{\left(x-\xi \right)U_0\left(\xi ,\tau \right){\exp \left(\frac{M}{a}\tau \right)}}{{\left[{\left(x-\xi \right)}^2+y^2\right]}^2}}}\times \nonumber \\
&\times \left[1-\mathrm{\Gamma }\left(2,\frac{{\left(x-\xi \right)}^2+y^2}{4\left(t-\tau \right)}\right)\right]d\xi d\tau +\frac{y}{\pi }\int^{\infty }_{-\infty }{\frac{T_0\left(\xi ,t\right)}{{\left(x-\xi \right)}^2+y^2}d\xi },
\end{align}
where $\mathrm{\Gamma }\left(\nu ,x\right)$ is the upper incomplete gamma function defined by $\mathrm{\Gamma }\left(\nu ,x\right)=\int^{\infty }_x{t^{\nu -1}{\exp \left(-t\right)}dt}$, and $\mathrm{\Gamma }\left(\nu ,0\right)=\mathrm{\Gamma }\left(\nu \right)$ is the gamma function. 

\noindent \textbf{\textit{Remark 3.1.}} In the derivation of integral representations \eqref{27}-\eqref{28}, the following relations have been utilized \cite{r39}:

\begin{enumerate}
\item  ${\mathcal{F}}^{-1}\left\{{\exp \left(-\left|q\right|y\right)}\right\}\left(x,y\right)=\frac{y}{\pi \left(x^2+y^2\right)}.$

\item  ${\mathcal{F}}^{-1}\left\{\frac{i}{q}{\exp \left(-\left|q\right|y\right)}\right\}\left(x,y\right)=-\frac{1}{\pi }{{\mathrm{tan}}^{-1} \left(\frac{x}{y}\right)}.$

\item  ${\mathcal{L}}^{-1}\left\{{\mathcal{F}}^{-1}\left[\frac{iq{\exp \left(-\left|q\right|y\right)}}{s+\frac{M}{a}}\right]\right\}\left(x,y,t\right)=-\frac{2xy{\exp \left(-\frac{M}{a}t\right)}}{\pi {\left(x^2+y^2\right)}^2}.$

\item  ${\mathcal{L}}^{-1}\left\{{\mathcal{F}}^{-1}\left[\frac{iq{\exp \left(-y\sqrt{q^2+s+\frac{M}{a}}\right)}}{s+\frac{M}{a}}\right]\right\}\left(x,y,t\right)=-\frac{2xy{\exp \left(-\frac{M}{a}t\right)}}{\pi {\left(x^2+y^2\right)}^2}\mathrm{\Gamma }\left(2,\frac{x^2+y^2}{4t}\right).$
\end{enumerate}

\noindent The Green's functions for the temperature and velocities in the semi-infinite domain can be obtained by setting \cite{r40}:

\begin{equation}\label{29}
 U_0\left(x,t\right)=-\delta \left(x\right)\delta \left(t\right),\ \ \ T_0\left(x,t\right)=\delta \left(x\right){\exp \left(-rt\right)},
\end{equation}
where $\delta \left(x\right)$ is the Dirac delta function and $r$ is a temporal heat dissipation coefficient. The first boundary condition, $U_0\left(x,t\right)$, stands for an instantaneous suction line along $z$-axis starts initially at $t=0$ as an instantaneous suction thrust with velocity equals one.  The presence of the exponential decay term in $T_0\left(x,t\right)$ is physically consistent and it means that the temperature resulting from the instantaneous suction decays exponentially and does not occur instantaneously. Therefore, the Green's functions for the stream function and the temperature can be then derived upon using \eqref{29} and \eqref{27}-\eqref{28}, we obtain

\begin{equation}\label{30}
\psi \left(x,y,t\right)=\frac{2xy{\exp \left(-\frac{M}{a}t\right)}}{{\pi \left(x^2+y^2\right)}^2}\left[1-\mathrm{\Gamma }\left(2,\frac{x^2+y^2}{4t}\right)\right]-\frac{1}{\pi }\delta \left(t\right){{\mathrm{tan}}^{-1} \left(\frac{x}{y}\right)},
\end{equation}
\begin{equation}\label{31}
\theta \left(x,y,t\right)=\frac{2k_1k_2xy{\exp \left(-\frac{M}{a}t\right)}}{{\pi a\left(x^2+y^2\right)}^2}\left[\mathrm{\Gamma }\left(2,\frac{x^2+y^2}{4t}\right)-1\right]+\frac{y{\exp \left(-rt\right)}}{\pi \left(x^2+y^2\right)}.
\end{equation}

Instead, let us consider two instantaneous suction lines at $x=\pm b$ which requires replacing \eqref{29} with 

\begin{align}\label{32}
U_0\left(x,t\right)=&-\frac{1}{2}\left[\delta \left(x+b\right)+\delta \left(x-b\right)\right]\delta \left(t\right),\nonumber \\ T_0\left(x,t\right)=&\frac{1}{2}\left[\delta \left(x+b\right)+\delta \left(x-b\right)\right]{\exp \left(-rt\right)}
\end{align}

The boundary conditions \eqref{32} with the integral representations \eqref{27}-\eqref{28} lead to the exact formulas

\begin{align}\label{33}
	\psi \left(x,y,t\right)=\frac{\left(x+b\right)y{\exp \left(-\frac{M}{a}t\right)}}{{\pi \left[{\left(x+b\right)}^2+y^2\right]}^2}\left[1-\mathrm{\Gamma }\left(2,\frac{{\left(x+b\right)}^2+y^2}{4t}\right)\right]-\frac{1}{2\pi }\delta \left(t\right){{\mathrm{tan}}^{-1} \left(\frac{x+b}{y}\right)} \nonumber \\
 +\frac{\left(x-b\right)y{\exp \left(-\frac{M}{a}t\right)}}{{\pi \left[{\left(x-b\right)}^2+y^2\right]}^2}\left[1-\mathrm{\Gamma }\left(2,\frac{{\left(x-b\right)}^2+y^2}{4t}\right)\right]-\frac{1}{2\pi }\delta \left(t\right){{\mathrm{tan}}^{-1} \left(\frac{x-b}{y}\right)},
\end{align}

\begin{align}\label{34}
	\theta \left(x,y,t\right)=\frac{k_1k_2\left(x+b\right)y{\exp \left(-\frac{M}{a}t\right)}}{{\pi a\left[{\left(x+b\right)}^2+y^2\right]}^2}\left[\mathrm{\Gamma }\left(2,\frac{{\left(x+b\right)}^2+y^2}{4t}\right)-\mathrm{1}\right]+\frac{y{\exp \left(-rt\right)}}{2\pi \left({\left(x+b\right)}^2+y^2\right)} \nonumber \\
 +\frac{k_1k_2\left(x-b\right)y{\exp \left(-\frac{M}{a}t\right)}}{{\pi a\left[{\left(x-b\right)}^2+y^2\right]}^2}\left[\mathrm{\Gamma }\left(2,\frac{{\left(x-b\right)}^2+y^2}{4t}\right)-\mathrm{1}\right]+\frac{y{\exp \left(-rt\right)}}{2\pi \left({\left(x-b\right)}^2+y^2\right)}.
\end{align}
The velocity components can be calculated from the stream function through the relations \eqref{17}.

\section{ Discussion}
\label{sec4}

This section is designated for discussing the exact formulas for the stream function \eqref{33}, the temperature \eqref{34}, and the resulting velocity components \eqref{17}. The mercury is chosen as the liquid metal, and the wall is arbitrarily chosen as a metal that makes a thermoelectric junction with mercury, see e.g., \cite{r41}. The temperature difference between the two hot lines $x=\pm b$ on the metallic wall and the contagious mercury (liquid metal) may generate an induced electromotive force, which known as ``Seebeck emf''. The induced electric currents may result a voltage difference on the boundary $y=0$, which in turn changes the temperature difference through cooling or heating. The latter is named ``Peltier effect''. The thermophysical properties of mercury Hg are given at room temperature $T_{\infty }=300{}^\circ \ \mathrm{K}$ as, see p. 1001 in \cite{r12}, 

\begin{align}\label{35}
\rho =13529\ & \mathrm{Kg}\cdot{\mathrm{m}}^{\mathrm{-}\mathrm{3}} \quad c_p=139.3\ \mathrm{J}\cdot \mathrm{K}{\mathrm{g}}^{\mathrm{-}\mathrm{1}}\cdot {\mathrm{K}}^{\mathrm{-}\mathrm{1}} \quad \nu =0.11\times {10}^{-6}\ {\mathrm{m}}^{\mathrm{2}}\cdot{\mathrm{s}}^{\mathrm{-}\mathrm{1}} \nonumber \\
&\kappa =8.54\ \mathrm{W}\cdot{\mathrm{m}}^{\mathrm{-}\mathrm{1}}\cdot{\mathrm{K}}^{\mathrm{-}\mathrm{1}}\quad p_r=0.0248 \quad {\sigma }_0={10}^6\ \mathrm{S}\cdot{\mathrm{m}}^{\mathrm{-}\mathrm{1}} 
\end{align}

Furthermore, we arbitrarily choose $r=1$, $b=0.5$, the Seebeck coefficient as $k_0=0.1\times {10}^{-3}\ \mathrm{V}\cdot{\mathrm{K}}^{\mathrm{-}\mathrm{1}}$, therefore the Peltier coefficient reads ${\pi }_0=30\times {10}^{-3}\ \mathrm{V}$, the characteristic velocity $V_0=1\ \mathrm{m}\cdot{\mathrm{s}}^{\mathrm{-}\mathrm{1}}$, $T_w=600{}^\circ \ \mathrm{K}$, and the magnetic induction $B_0=10\ \mathrm{Tesla}$, hence the figure of merit and the other parameters of the problem can be set as

\begin{equation}\label{36}
M=8.13\times {10}^{-4} \quad ZT_{\infty }=0.35 \quad k_1=22.17 \quad k_2=5.8\times {10}^{-7}.
\end{equation}

\begin{figure}
	\centering
	\includegraphics*[scale=0.5,angle=0]{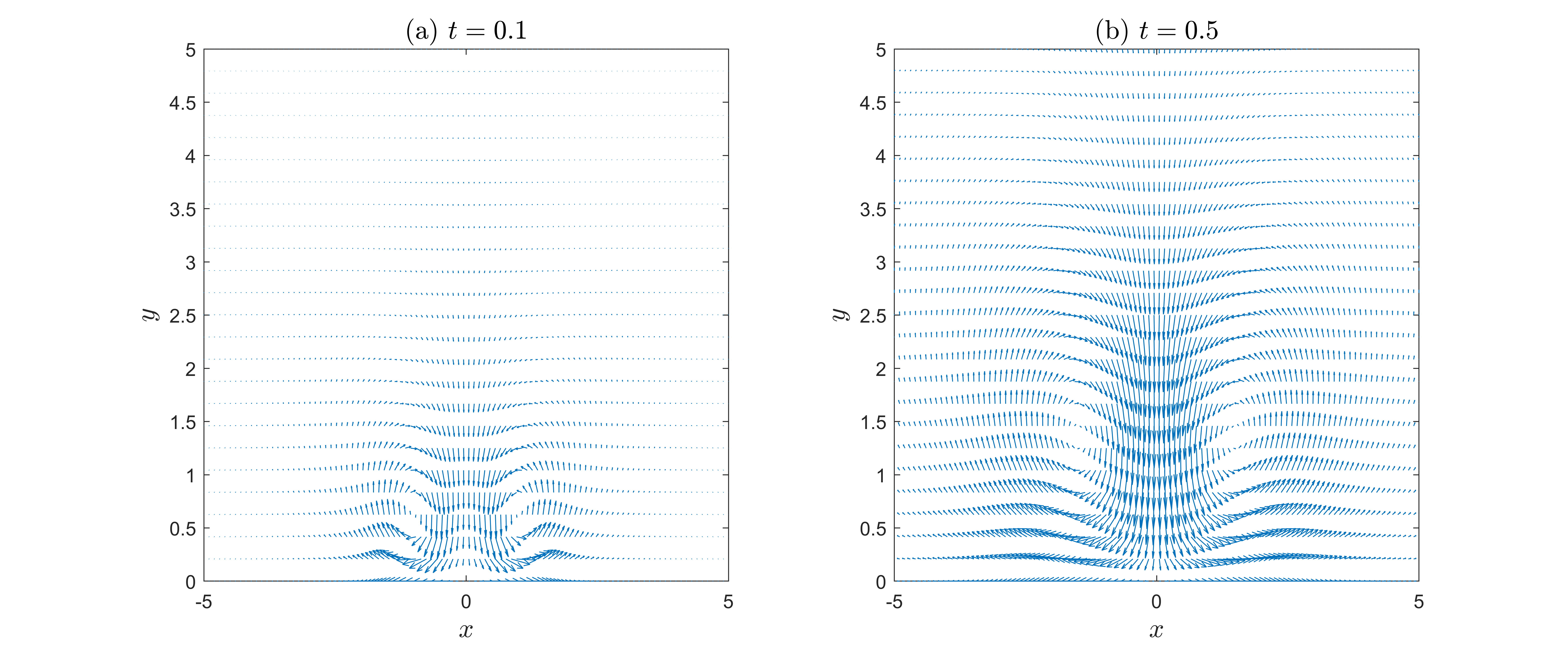}
	\includegraphics*[scale=0.5,angle=0]{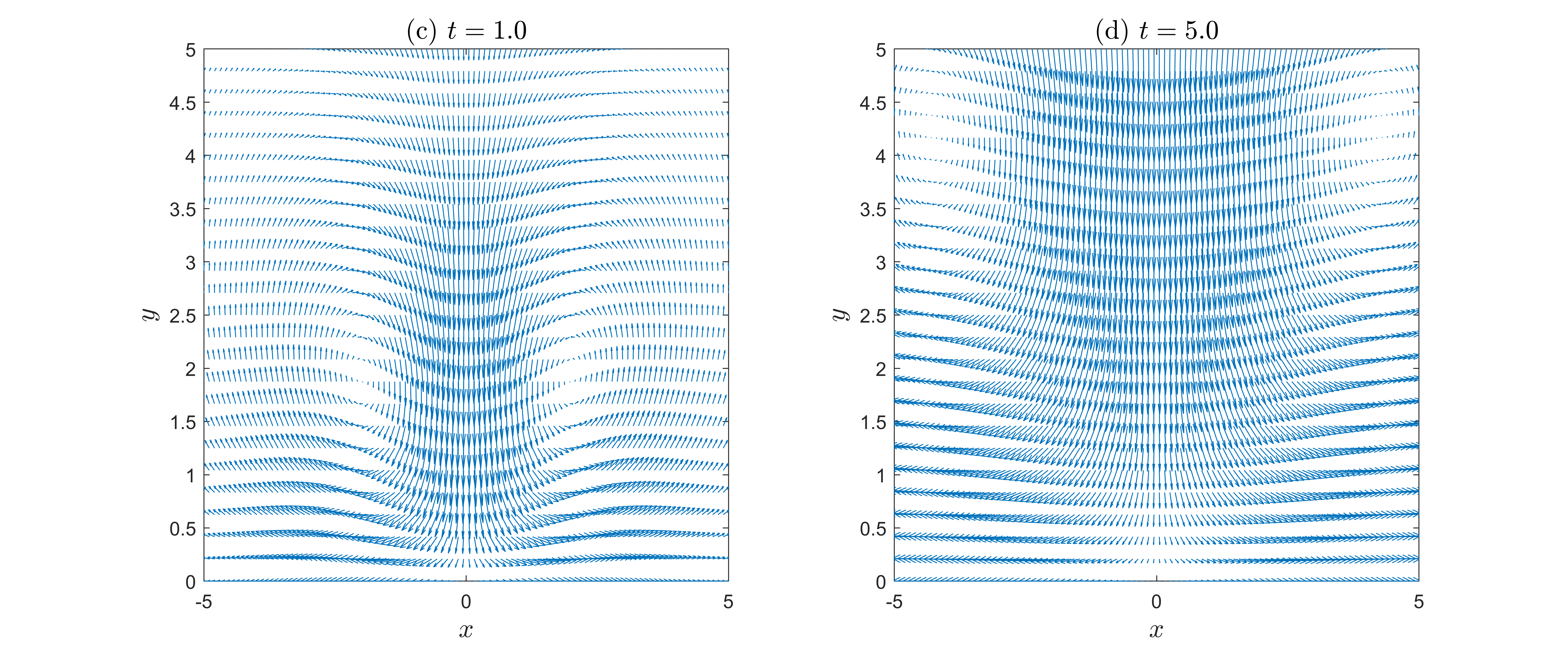}
	\caption{Velocity vector in \textit{$xy$}-plane at different values of time.}
	\label{fig2} 
\end{figure}

\begin{figure}
	\centering
	\includegraphics*[scale=0.5,angle=0]{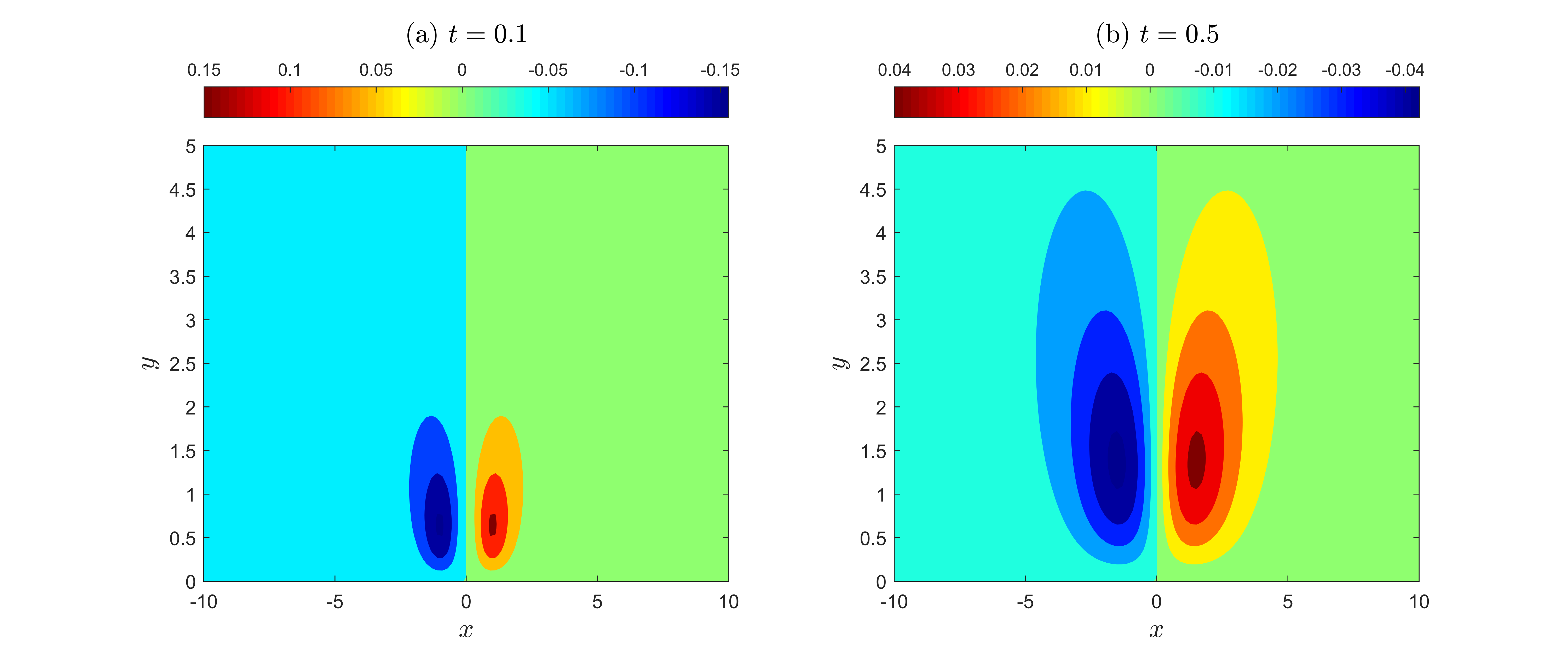}
	\caption{Stream function at two instants \textit{$t=0.1$} and \textit{$t=0.5$}.}
	\label{fig3} 
\end{figure}

\begin{figure}
	\centering
	\includegraphics*[scale=0.5,angle=0]{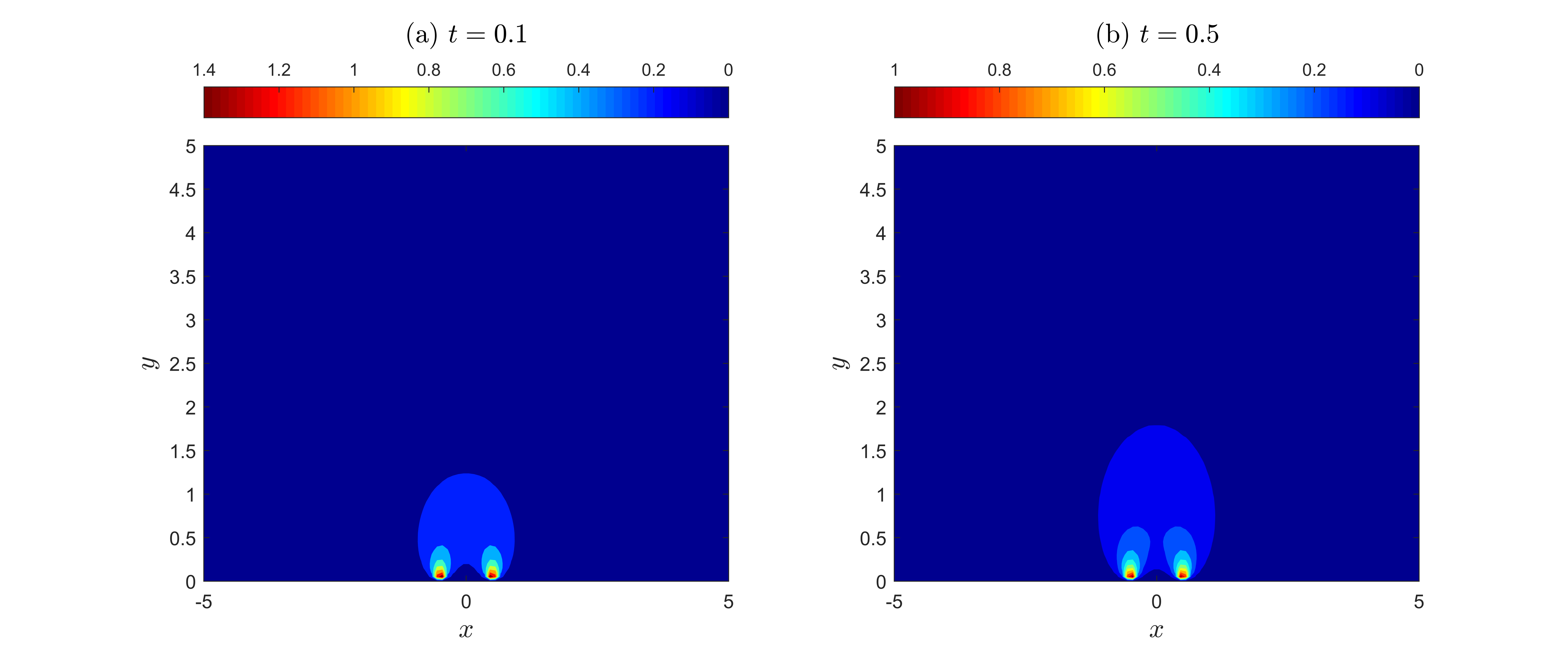}
	\includegraphics*[scale=0.5,angle=0]{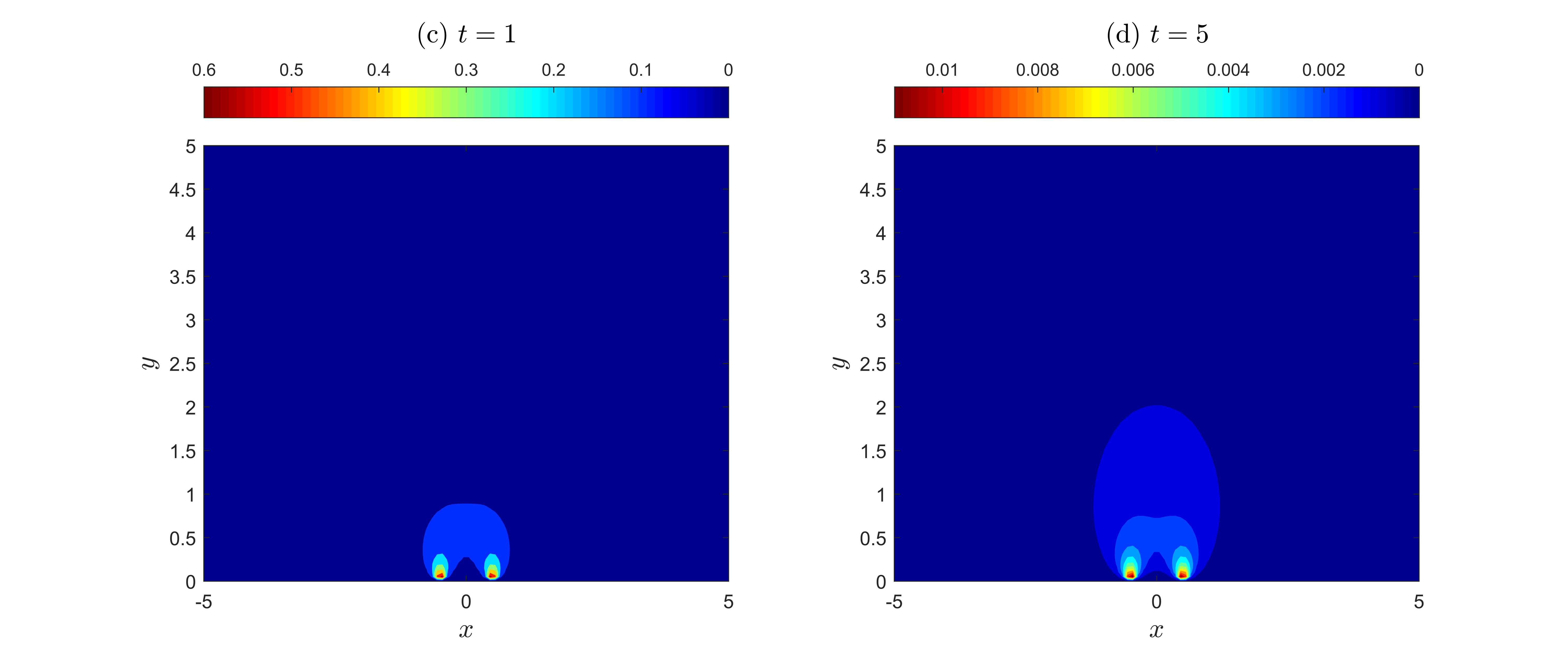}
	\caption{Temperature distribution at different values of time.}
	\label{fig4} 
\end{figure}

\begin{figure}
	\centering
	\includegraphics*[scale=0.5,angle=0]{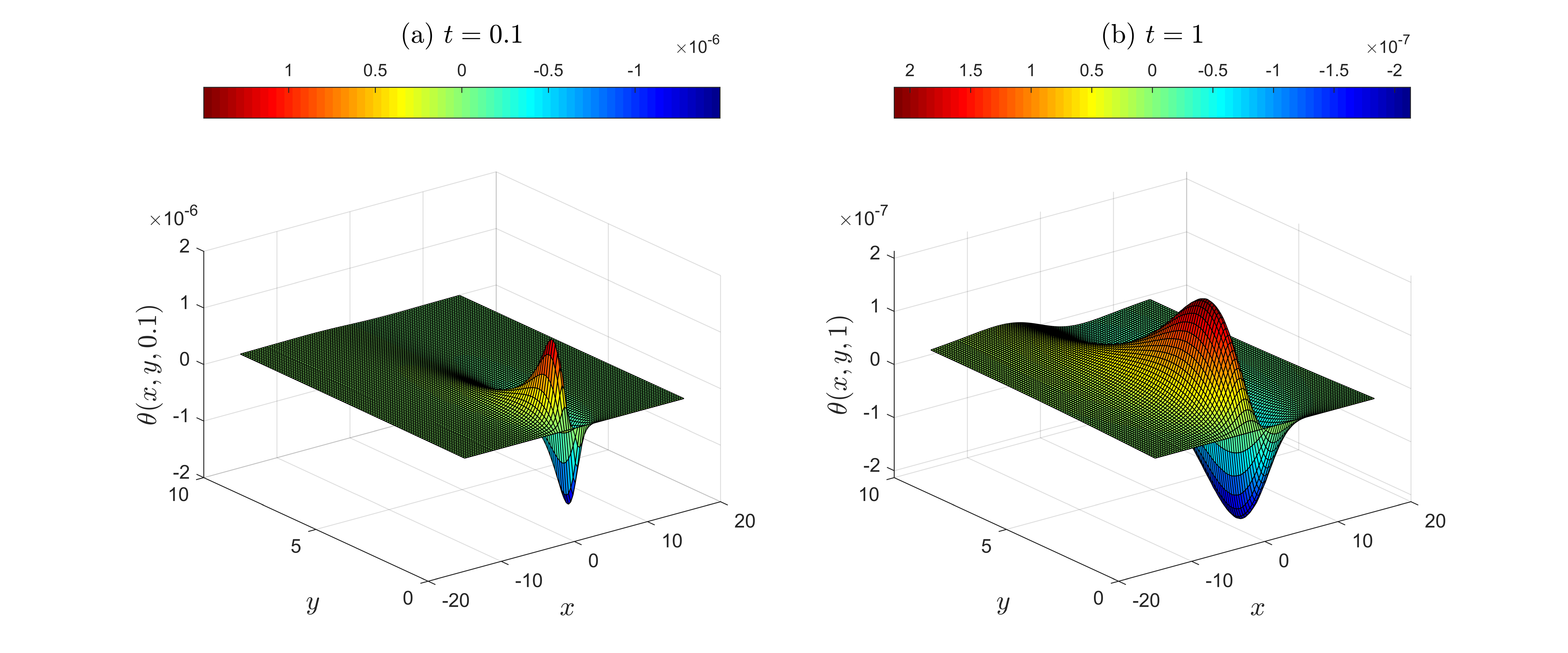}
	\caption{ Induced temperature for the case $r\to \infty $.}
	\label{fig5} 
\end{figure}

We present a depiction of the liquid velocity vector, Eqs \eqref{17} and \eqref{33}, at different values of time in  Fig. \ref{fig2},  while the stream function is graphically represented in Fig. \ref{fig3}. The temperature of the liquid metal, Eq. \eqref{34}, is shown in Fig. \ref{fig4}. One can easily note the slow variation of temperature with time. This feature is a natural consequence of the steady state heat transfer assumption. It is clear from equation \eqref{34} that the temperature of thermoelectric liquid comes from two main parts: the first is the exponentially decaying temperature imposed on the two instantaneous suction lines, $x=\pm b$, and the second part is the temperature produced from the Peltier effect. Therefore, the variation of temperature in Fig. \ref{fig4} is mainly resulting from the thermal boundary condition, more precisely, ${\exp \left(-rt\right)}$.  In the case $r=0$, the liquid temperature of the two lines $x=\pm b$ on the wall is kept constant during motion, therefore, the only temporal variation for the temperature is due to the thermoelectric term characterized by the exponential ${\exp \left(-Mt/a\right)}$, refer to \eqref{31} and \eqref{34}. In this case, the temporal variation of temperature due to thermoelectric currents is neglected when compared to the constant non-zero temperature at $x=\pm b$, namely, $y/\left[\pi \left(x^2+y^2\right)\right]$. On the other hand, the case $r\to \infty $ in which the temperature resulting from instantaneous suction velocity is neglected, i.e., $\theta \left(x,0,t\right)=0,\ \ \ t\ge 0$, yields a thermal couple due solely to the thermoelectric currents, refer to the first term of \eqref{31}. In this case the temporal variation of temperature is recovered, and two opposite regions are arisen, one has a negative temperature and the other has a positive temperature, see Fig. \ref{fig5}. 

The thermoelectric effects on the velocity magnitude $\left|{\mathbf{v}}\right|=V=\sqrt{u^2+v^2}$ is shown in Table \ref{table1}. Apparently, the thermoelectric effects increase slightly the velocity magnitudes at all positions. \newline
\noindent 
\begin{table}
\begin{tabular}{|p{0.75in}|p{1.21in}|p{1.21in}|p{1.21in}|p{1.21in}|} \hline 
& \multicolumn{2}{|p{2.0in}|}{$x=0$} & \multicolumn{2}{|p{2.0in}|}{$x=0.5$} \\ \hline 
$y$ & 0.01 & 1 & 0.01 & 1 \\ \hline 
$k_1=22.17$ & 0.002808142983056 & 0.1366381519467 & 0.226849951203674 & 0.175568864493767 \\ \hline 
$k=0$ & 0.002808083794266 & 0.1366352719485 & 0.226845169761204 & 0.175565163931254 \\ \hline 
\end{tabular}
\caption{Velocity magnitudes at different positions and dimensional time \textit{$t=0.1$} with and without thermoelectric effects.}
\label{table1}
\end{table}

\section{ Conclusions}
\label{sec5}

In a summary, we have presented a two-dimensional formulation for the thermoelectric liquids on a vertical plate with partial suction area. The heat transfer has been assumed to be steady. The steady-state heat flow assumption is reasonable in the case of liquid metal because its Prandtl number is very small. The slow motion has been assumed, as such, the problem is linearized and becoming more tractable from an analytical point of view. Lastly, we have neglected the magnetic field distortion. The Green's functions have been analytically derived for the temperature and the stream function.  Exact forms for the temperature and the stream function have been derived for the case of two instantaneous hot suction lines on the vertical plate. Two opposite vortices are arisen inside the liquid and expanding with time. The temperature variation is slow with time because of the steady-state assumption. The main cause of temporal variation of temperature is the exponential decaying surface temperature with rate $r$. Two main cases have been discussed: First, $r=0$, and second, $r\to \infty $. In the case $r=0$, the temporal variation of temperature is neglected, while the case $r\to \infty $ yields temperature profile resulting from the thermoelectric currents. Finally, the effects of thermoelectricity on the velocity are studied and we find very small increments in the velocity magnitudes. In a future work, a numerical investigation is under preparation, where it will address the unsteady heat flow, in addition to the distorted magnetic field.

\section*{Acknowledgments}
E.A. is grateful to Prof Magdy Ezzat for fruitful discussion about his works in TEMHD.

\end{document}